\documentclass[aps,prl,superscriptaddress,onecolumn,oneside,floatfix,nofootinbib,preprint]{revtex4}

\usepackage{amsmath}
\usepackage{amsfonts}
\usepackage{graphicx}
\usepackage{mathrsfs}
\usepackage{rotating}
\usepackage{setspace}
\usepackage{supertabular}
\usepackage{amsthm}
\usepackage{subfigure}
\usepackage{multirow}

\begin{document}
\title{Nonequilibrium static diverging length scales on approaching a prototypical model glassy state}

\author{Adam B. Hopkins}
\affiliation{Department of Chemistry, Princeton University, Princeton, New
Jersey 08544, USA}
\author{Frank H. Stillinger}
\affiliation{Department of Chemistry, Princeton University, Princeton, New
Jersey 08544, USA}
\author{Salvatore Torquato}
\affiliation{Department of Chemistry, Princeton University, Princeton, New
Jersey 08544, USA}
\affiliation{Princeton Institute for the Science and Technology of Materials,
Princeton University, Princeton, New Jersey 08544, USA}
\affiliation{Department of Physics, Princeton University, Princeton, New Jersey
08544, USA}
\affiliation{Princeton Center for Theoretical Science, Princeton University,
Princeton, New Jersey, 08544, USA}
\affiliation{Program in Applied and Computational Mathematics, Princeton
University, Princeton, New Jersey 08544, USA}

\begin{abstract}
Maximally random jammed states of hard spheres are prototypical glasses. We study the small wavenumber $k$ behavior of the structure factor $S(k)$ of overcompressed million-sphere packings as a function of density up to the jammed state. We find both a precursor to the glassy jammed state evident long before the jamming density is reached and an associated growing length scale, extracted from the volume integral of the direct correlation function $c(r)$, which diverges at the ``critical'' jammed state. We also define a nonequilibrium index $X$ and use it to demonstrate that the packings studied are intrinsically nonequilibrium in nature well before the critical point is reached.
\end{abstract}
% 68.08.De Liquid-solid interface structure: measurements and simulations
% 61.46.Bc Structure of clusters
% 82.60.Nh Thermodynamics of nucleation
% 61.43.-j Disordered solids

\maketitle

A sufficiently rapid quench of a liquid from above its freezing temperature into a supercooled regime can avoid crystal nucleation to produce a glass with a relaxation time that is much larger than experimental time scales, resulting in an amorphous state (without long-range order) that is simultaneously rigid \cite{CLPCMP1995}. The underlying physics of the glass transition is one of the most fascinating open questions in materials science and condensed-matter physics. Many conundrums remain, including whether the growing relaxation times under supercooling have accompanying growing structural length scales. Essentially, two opposing explanations have emerged to address this question. One asserts that a static structural length scale does not exist and identifies growing dynamical length scales \cite{BBBKMR2007a,KDS2009a,CG2010a}. The other contends that there is a static growing length scale of thermodynamic origin \cite{LW2006a,HMR2012a}. In this paper, we present both theoretical and computational results that support an alternative view, namely, the existence of a growing static length scale but one that is intrinsically nonequilibrium in nature.

Our model systems are disordered packings of identical spheres with densities between the freezing transition and the so-called maximally random jammed (MRJ) state \cite{TTD2000a}. The MRJ state under the strict-jamming constraint is a {\it prototypical} glass in that it lacks any long-range order but is perfectly rigid (the elastic moduli are indeed unbounded) \cite{TS2007a,TS2010a}. This endows such packings with special attributes. For example, MRJ packings are hyperuniform \cite{TS2003a,DST2005a} ({\it i.e.}, infinite wavelength density fluctuations vanish) with a structure factor $S(k)$ that tends to zero linearly in the wavenumber $k$, implying quasi-long-ranged negative pair correlations (anticorrelations) decaying as a power law \cite{DST2005a}. This large-scale property is markedly different from typical liquids in equilibrium, which tend to exhibit more rapidly decaying pair correlations (including exponential decays).

It has been theoretically shown that hyperuniform point distributions are at an ``inverted'' critical point, {\it i.e.}, in contrast to normal fluid critical points, the direct correlation function $c(r)$, rather than the total correlation function $h(r) \equiv g_2(r) -1$, with $g_2(r)$ the pair correlation function, becomes long-ranged (decaying more slowly than $1/r^d$, where $d$ is the space dimension and $r$ is the radial distance) \cite{TS2003a}. This suggests that the volume integral of $c(r)$ for overcompressed hard-sphere configurations \cite{endnote1} that follow Newtonian dynamics should grow as the MRJ state is approached and ultimately diverge at this {\it inverted} critical point. In this Letter, we show that this not only is the case but that the nonequilibrium signature of quasi-long-range anticorrelations, which we quantify via a {\it nonequilibrium index} $X$, emerges well before the jammed state is reached. This implies that the direct correlation function of a glass formed by supercooling a liquid (in which the molecules possess both repulsive and attractive interactions) provides a static growing length as its glass (jamming) transition is approached. Hence, our findings, based on nonequilibrium hard-sphere systems, could be applied more broadly to general glass formers, as we will elaborate below.

The present study focuses on prototypical glassy states represented by three-dimensional identical nonoverlapping spheres. For this system, it has been shown that MRJ packings are characterized by quasi-long-ranged pair anticorrelations in which $h(r)$ decays as $-1/r^4$ \cite{DST2005a}. This quasi-long-ranged behavior in $r$ is equivalent to linear behavior in the structure factor $S(k)$ near the origin with $S(k)$ nonanalytic at $k=0$. The structure factor $S(k)$ is defined in terms of the Fourier transform of $h(r)$, $S(k) = 1 + \rho\tilde{h}(k)$, with $\rho$ the number density, equal to $(6/\pi)\phi$ for unit-diameter hard spheres with packing fraction $\phi$ the fraction of space covered. Quasi-long-ranged anticorrelations have been shown to be present in the ground states of liquid helium \cite{RC1967a} and noninteracting spin-polarized fermions \cite{TSZ2008a}, in the Harrison-Zeldovich spectrum of the early Universe \cite{PeeblesPC1993}, and in MRJ states of Platonic solids \cite{JT2011a}. Experimentally, the most accurate measurements of the structure factor of bulk amorphous silicon have revealed a linear trend in the small-$k$ behavior of $S(k)$ toward $S(0) = 0$ that appears to be consistent with hyperuniformity \cite{LXWMRTS2012a}. Linear behavior near the origin in $S(k)$ for a hyperuniform system indicates an inverted critical point, characterized by long-range anticorrelations in a direct correlation function $c(r)$ that decays as $-1/r^2$, where we define $c(r)$ in terms of the Fourier transform of $\tilde{c}(k)$,
\begin{equation}
\tilde{c}(k) = \frac{\tilde{h}(k)}{S(k)} = \frac{S(k)-1}{\rho S(k)}.
\label{CK}
\end{equation}
More information about this inverted critical point, including critical exponents, can be found in Ref. \cite{TS2003a}.

\begin{figure}[ht] 
\centering
\includegraphics[angle=270,width = 6.0in,viewport = 80 25 585 725,clip]{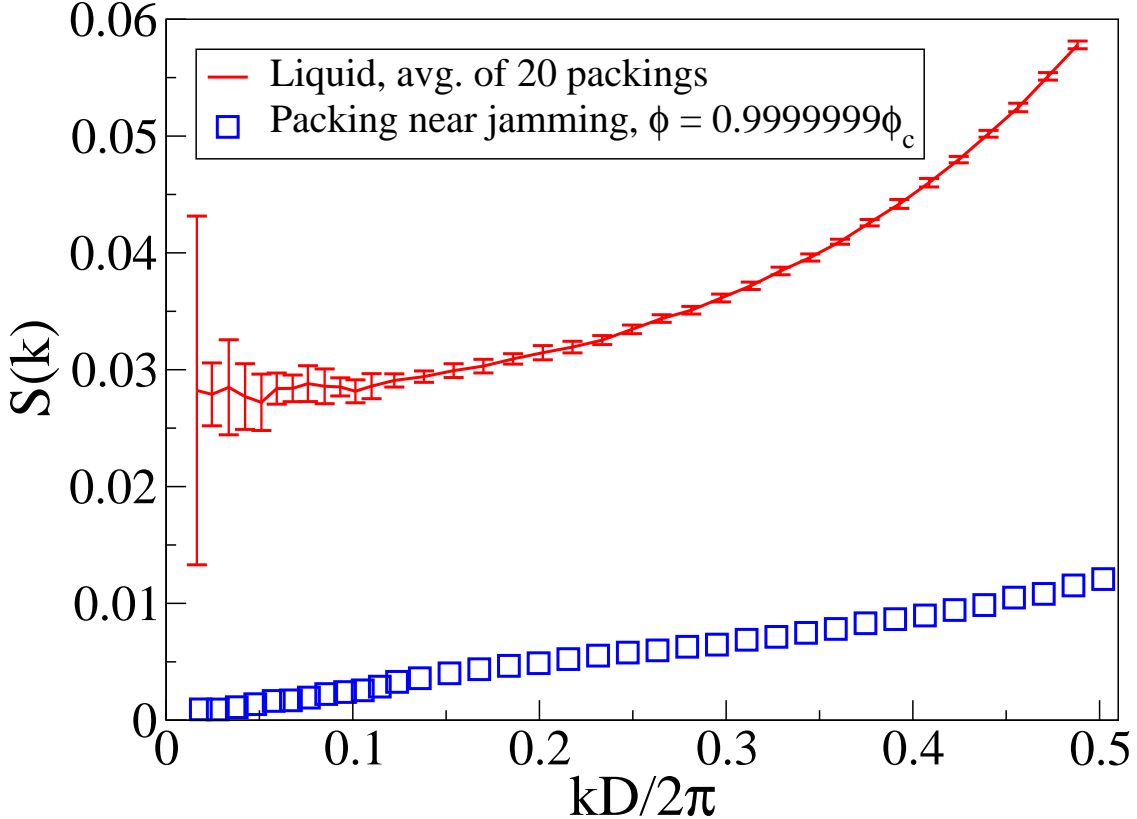}
\caption{(color online). Bottom curve: angularly-averaged $S(k)$ calculated by DFT for a million-sphere packing with $\phi=0.9999999\phi_c\dots$. Note the linear behavior in $S(k)$ as $k\rightarrow 0$. Top curve: average of $20$ angularly-averaged $S(k)$ calculated by DFT for independently generated million-sphere packings in the equilibrium liquid state with $\phi = 0.45$. Vertical bars represent an estimate of the standard deviation of the distribution of values for $S(k)$ at each $k$ over the $20$ packings.}
\label{introFig}
\end{figure}

In this work, we study million-particle packings of identical spheres overcompressed at various rates fast enough to avoid the formation of crystallites. We employ an event driven friction-free Lubachevsky-Stillinger molecular dynamics protocol under periodic boundary conditions \cite{DTS2005a} in a cube of side length unity where initial sphere velocities are Maxwell-Boltzmann distributed such that mean energy per sphere (of mass unity) is $k_BT = 1/2$, with $k_B$ Boltzmann's constant and $T$ temperature. Simulations are performed to overcompress spheres from various initial states (including equilibrium liquid) at densities below freezing $\phi_f = 0.494$ up to various percentages of the jamming fraction $\phi_c$. For each packing, $\phi_c$ is dependent upon initial conditions and compression rate, but it varies no more than $5\times 10^{-5}$ for packings compressed at the same rate due to the very large system size. We study rates in this work corresponding to sphere diameter growth per unit time from $\Gamma = 0.0007$ to $\Gamma = 0.03$ \cite{DTS2005a}, where $\Gamma = 0.0007$ appears to be the slowest rate at which crystallites do not form.

We closely examine the small wavenumber $k$ behavior of the angularly-averaged structure factor $S(k)$, where $k$ is the magnitude of a wavevector ${\bf k}$. We calculate $S(k)$ by direct Fourier transform (DFT), omitting forward scattering, for $N = 1,000,000$ unit-diameter spheres in a periodic cube of length $L$ replicated over all space, using $S({\bf k}) = (1/N)|\sum_{j=1}^{N}e^{i{\bf k} \cdot {\bf r}_j}|^2$, and angularly averaging over all ${\bf k}$ of equal magnitude $k$. The smallest nonzero wavenumber $k$ calculated by this method is equal to $2\pi/L$, which for $1,000,000$ unit-diameter spheres at jamming is about $0.067$. Figure \ref{introFig} compares the behavior of $S(k)$ for a) a single packing overcompressed from the equilibrium liquid at $\phi = 0.69\phi_c$ to $\phi = 0.9999999\phi_c$ and b) an average of $20$ independently generated packings in the equilibrium liquid state with $\phi=0.45\sim 0.70\phi_c$.

\begin{figure}[ht] 
\centering
\includegraphics[angle=270,width = 6.0in,viewport = 45 10 600 750,clip]{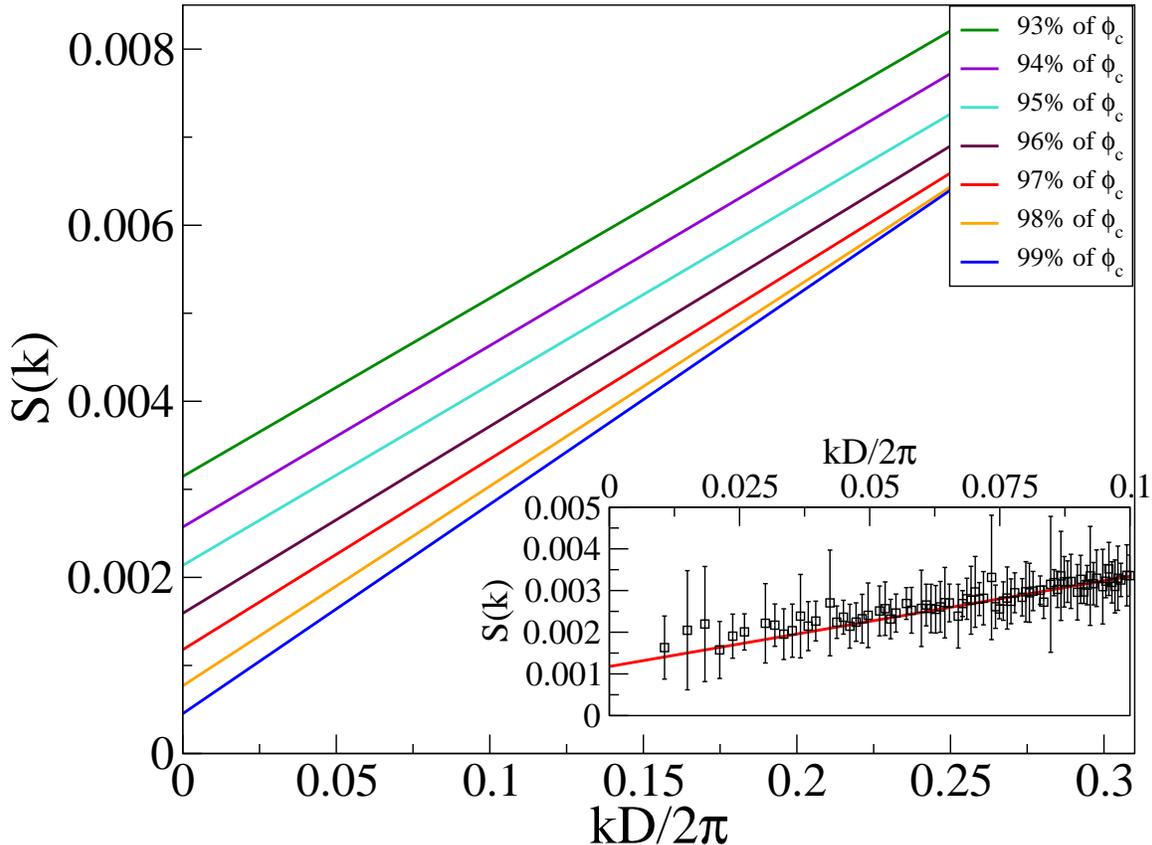}
\caption{(color online). Linear fits with $\phi$ at various percentages of jamming fraction $\phi_c$ to the average of $20$ angularly-averaged structure factors $S(k)$ calculated by DFT from million-sphere packings compressed at a rate $\Gamma = 0.03$ from the equilibrium liquid. All fits achieve R-squared values of greater than $0.995$. The curves in the figure from top to bottom correspond to the labels in the legend from top to bottom. Inset: average data and estimate of the standard deviation (over the $20$ packings) for angularly-averaged $S(k)$ for packings compressed to $\phi=0.97\phi_c$.}
\label{SKfits}
\end{figure}

For all compression rates and initial conditions studied, we identify a precursor to quasi-long-ranged pair anticorrelations manifest in linear behavior in $S(k)$, {\it i.e.}, $S(k) = ak + b$. For $\phi$ greater than about $0.92\phi_c$, this behavior extends from the smallest $k$ studied to about $k\sim 0.4$, whereas for $\phi_f < \phi < 0.93\phi_c$, a linear trend is clearly evident but does not extend to the smallest values of $k$ for all compression rates studied. This behavior is illustrated in Fig. \ref{SKfits} for a wide range of densities \cite{endnote3}.

Though plots of $S(k)$ from individual packings show variability at small $k$ around the linear trend attributable to the dynamics of particle movements as the spheres are compressed, the average of several packings always illustrates a distinctly linear trend. This trend additionally appears for packings compressed at various rates from a random sequential addition (RSA) state \cite{TorquatoRHM2002}, and the trend appears for packings compressed at fast but changing rates. Taken together, these observations indicate that the linear trend is easily reproducible across different compression rates and initial conditions, even nonequilibrium conditions like RSA, so long as the compression rate is fast enough such that crystallites do not form.

\begin{figure}[ht] 
\centering
\includegraphics[angle=270,width = 6.0in,viewport = 50 10 595 750,clip]{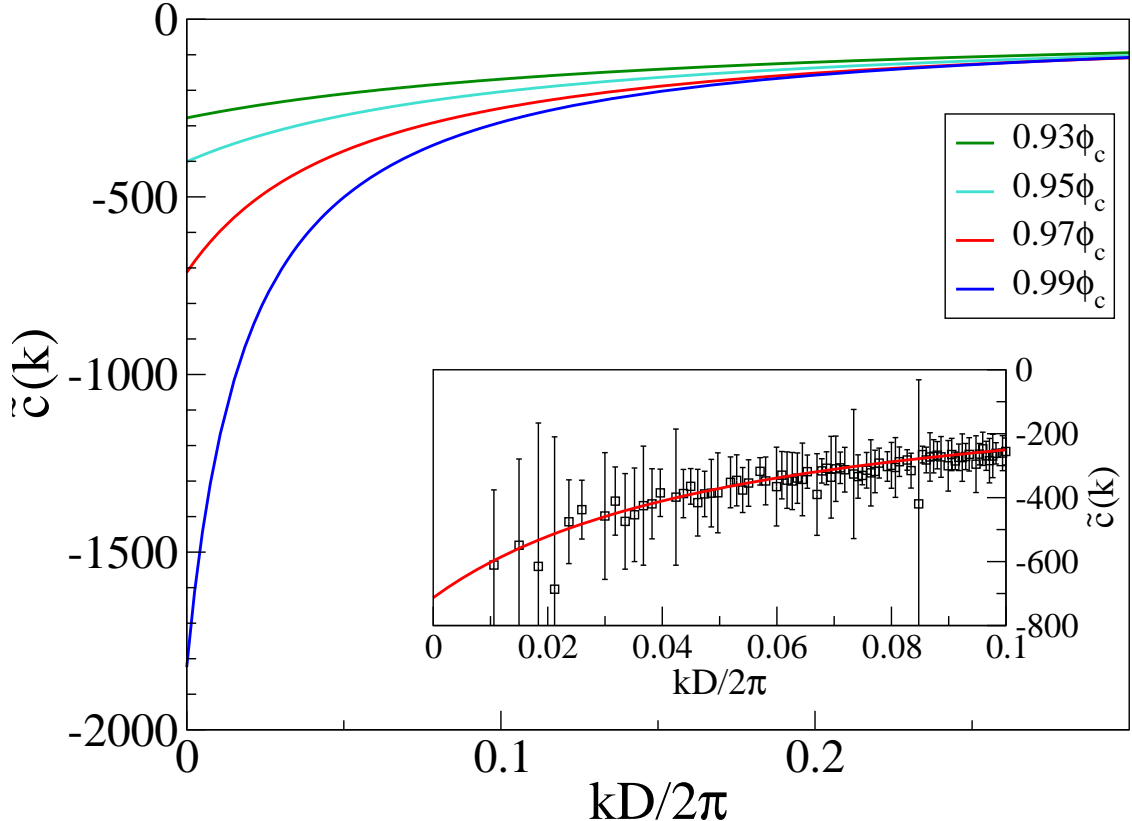}
\caption{(color online). Fits of angularly-averaged $\tilde{c}(k)$ calculated from the fits in Fig. \ref{SKfits}. Note that $\tilde{c}(0)$ appears to diverge to $-\infty$ as $\phi\rightarrow\phi_c$. Inset: average data and estimate of the standard deviation (over $20$ packings) for $\tilde{c}(k)$ for packings overcompressed to $\phi=0.97\phi_c$. For this data set, geometric ($\mu_g = (\prod_{i=1}^{20}x_i)^{1/20}$, with $\mu_g$ the mean and $x_i$ the data points) and arithmetic averages were nearly identical.}
\label{CKfits}
\end{figure}

As $\phi \rightarrow \phi_c$, the the values of $\tilde{c}(k)$ (Eq. (\ref{CK})) near $k=0$ begin to diverge. In particular, in the limit as $\phi\rightarrow\phi_c$ for a hyperuniform packing, $\tilde{c}(0)\rightarrow -\infty$. This divergent behavior can clearly be seen in Fig. \ref{CKfits}, which are plots of $\tilde{c}(k)$ calculated using Eq. (\ref{CK}) and the fits in Fig. \ref{SKfits}. 

The length scale $\xi_{DCF} \equiv (-\tilde{c}(0))^{1/3}$ grows continuously with packing fraction and diverges as the hyperuniform state is approached, indicating a long-ranged direct correlation function $c(r)$ that decays asymptotically proportional to $-1/r^2$. The growth in $\xi_{DCF}$ signals the incipient rigidity associated with the formation of nearly jammed sphere-contact networks, and $\xi_{DCF}$ diverges when the packing becomes completely rigid at the inverted critical point $\phi = \phi_c$. The presence of this inverted critical point suggests that renormalization group methods might be fruitfully applied to study the behavior of $c(r)$ in this and similar glassy systems.

In the mean spherical approximations \cite{HMTSL2006}, $c(r)$ can be thought of as the negative of an effective pair potential $v_{eff}(r)$, that is to say, $-\beta v_{eff}(r) = c(r)$ with $\beta = 1/k_BT$, which is asymptotically exact as $r\rightarrow \infty$. Following this interpretation, MRJ packings exhibit long-ranged repulsions $v_{eff}(r)$ that are asymptotically proportional to $1/r^2$, which can be thought of as a generalized Coulombic interaction $1/r^n$ with $n=2$ instead of $n=1$, that drive the system to have no infinite wavelength density fluctuations, {\it i.e.}, to be hyperuniform. In comparison, one-component plasmas in equilibrium interacting with a Coulomb potential (e.g., $v(r) \sim 1/r$ in 3D) are hyperuniform \cite{TS2003a,TSZ2008a,LWL1999a}.

Other diverging length scales can also be identified, {\it e.g.}, the inverse of the first point $k$ in $S(k)$ where linear behavior becomes dominant, which we term $\xi_{SF}$. This length scale is visually evident in $S(k)$ for the systems studied beginning at $\phi$ just above $\phi_f$, and it can be shown \cite{HST2012a} to correspond roughly to the point in sphere packings where $-1/r^4$ pair anticorrelations in $h(r)$ cease to be dominant and are supplanted by faster decay. This length scale $\xi_{SF}$ grows far more quickly than $\xi_{DCF}$; in particular, it is already greater than the size $L$ of the million-sphere systems studied once $\phi = 0.93\phi_c$.

The identification of static growing length scales that can be extracted from pair information is a novel and unexpected finding. Though diverging time scales have long been evident in glasses, quickly growing length scales have been difficult to find. Recent studies have highlighted the presence of growing length scales using both static and dynamic four-point correlation functions; however, none of these length scales have been shown to grow very large, and it has been suggested that standard pair correlation functions may lack the information necessary to describe the static growing length scales present in glasses \cite{BK2011a,CCT2012a}. Our results demonstrate otherwise. By refocusing investigations to examining the small-$k$ behaviors of $\tilde{c}(k)$ and $S(k)$, which are accessible to us via very large million-sphere packings, we are able to determine the large-$r$ behavior of $c(r)$ and $h(r)$ (which is difficult to do in real space) to extract growing length scales.

Though there are diverging length and time scales evident in the structure factors of these prototypical glasses, the packings are far from equilibrium. In particular, the compressibility equation relating the isothermal compressibility $\kappa_T = (1/\phi)(dp/d\phi)_T$, with $p$ the pressure, to infinite wavelength density fluctuations,
\begin{equation}
S(0) = \rho k_BT\kappa_T,
\label{compressibility}
\end{equation}
which holds for systems in {\it thermal equilibrium}, does not hold for the systems studied. To investigate this concept further, we make use of an expression for the reduced pressure $p/\rho k_BT$ that is valid continuously from the freezing point $\phi_f$ all the way up to the jamming density $\phi_c$ as obtained from the nearest-neighbor conditional pair distribution function $G(r)$ \cite{TorquatoRHM2002}. Specifically, \begin{equation}
G(\infty) = p/\rho k_BT = 1 + 4\phi g_f(1)\frac{\phi_c - \phi_f}{\phi_c - \phi},
\label{Ginfinity}
\end{equation}
where $g_f(1)$ is the value of $g_2(1)$ for an equilibrium liquid at $\phi_f$. From Eq. (\ref{Ginfinity}), we can derive the right hand side of Eq. (\ref{compressibility}),
\begin{equation}
\rho k_BT\kappa_T = \frac{(\phi_c - \phi)^2}{4\phi\phi_cg_f(1)(\phi_c-\phi_f)}.
\label{rhoKBTKT}
\end{equation}

\begin{figure}[ht] 
\centering
\includegraphics[angle=270,width = 6.0in,viewport = 50 0 595 750,clip]{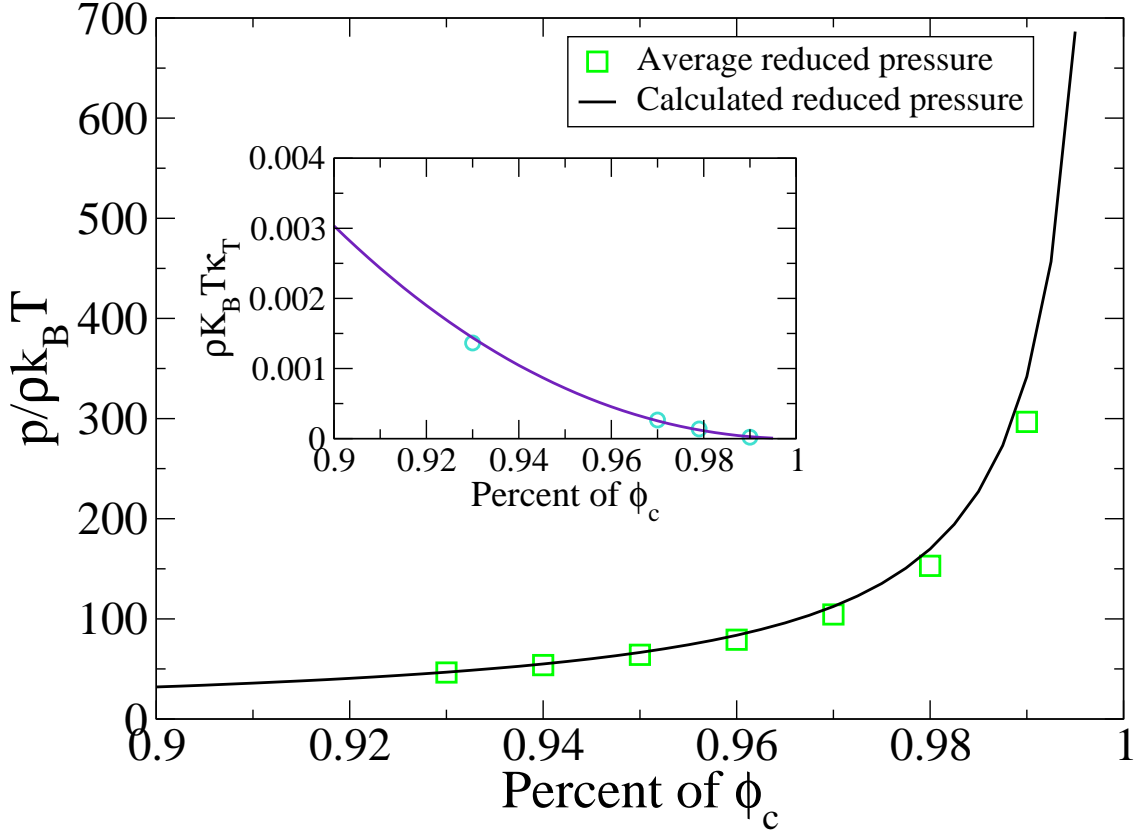}
\caption{(color online). Average over $20$ packings compressed at rate $\Gamma = 0.03$ of reduced pressure at various percentages of $\phi_c$, compared to the calculated values of reduced pressure from Eq. (\ref{Ginfinity}). Inset: average values of $\rho k_BT\kappa_T$ at each value of $\phi$ for four packings, compared to the calculated values using Eq. (\ref{rhoKBTKT}). The standard deviation of the measurement is about the size of the circle displayed for $\phi = 0.93\phi_c$ and much smaller than the circles for the other three measurements.}
\label{rhoKBTKTfits}
\end{figure}

Relation (\ref{Ginfinity}) has been shown to fit closely to data \cite{RT1998a}, and our results, displayed in Fig. \ref{rhoKBTKTfits}, support this finding. Values of $\rho k_BT\kappa_T$ calculated from Eq. (\ref{rhoKBTKT}) also fit the data. In Fig. \ref{rhoKBTKT}, we calculate $\kappa_T$ by rescaling velocities after each particle has undergone two collisions so that temperature remains constant. The expression $\rho k_BT\kappa_T$ is computed once every time particles have undergone $20$ collisions each from the relation $\rho k_BT\kappa_T = (1/\phi)(\Delta\phi/\Delta(p/\rho k_BT))$, with $p$ the pressure, and the average over ten calculations is taken. The average of these values over five packings is reported as the measured isothermal compressibility. Though reduced pressure, and therefore $\kappa_T$, should change if spheres are allowed to relax (collide with no compression), in our simulations this change was undetectable within the error even for values of $\rho k_BT\kappa_T$ after relaxation times of $100,000$ collisions per sphere at $\phi = 0.93\phi_c$.

\begin{figure}[ht] 
\centering
\includegraphics[angle=270,width = 6.0in,viewport = 50 10 595 760,clip]{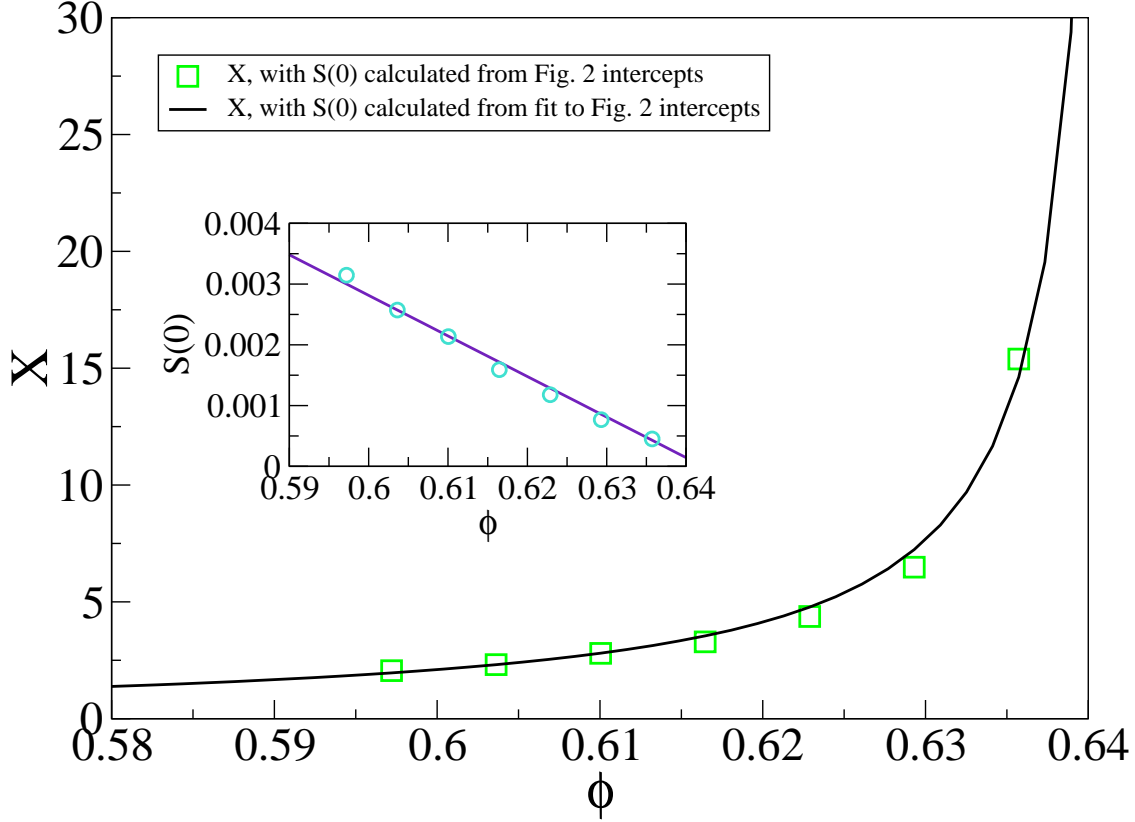}
\caption{(color online). Plot of $X$ with $\rho k_BT\kappa_T$ calculated from Eq. (\ref{rhoKBTKT}) and $S(0)$ taken from the linear fits in Fig. \ref{SKfits}, compared to where $S(0)$ is calculated from a linear fit to the intercepts $S(0)$ in Fig. \ref{SKfits}. Inset: linear fit to the intercepts $S(0)$ in Fig. \ref{SKfits}.}
\label{X}
\end{figure}

With Eq. (\ref{rhoKBTKT}) in mind, we define
\begin{equation}
X \equiv \frac{S(0)}{\rho k_BT\kappa_T} -1
\label{noneqIndex}
\end{equation}
as a {\it nonequilibrium index} to quantify the degree to which the systems under study deviate from thermal equilibrium $X=0$ [cf. Eq. (\ref{compressibility})]. Figure \ref{X} shows a plot of $X$ vs. $\phi$, demonstrating that $X$ diverges as $\phi\rightarrow\phi_c$.

Using Eq. (\ref{rhoKBTKT}) and fitting a linear trend to values of $S(0)$ with $\phi$ from Fig. \ref{SKfits} (inset of Fig. \ref{X}), the behavior of $X$ as $\phi \rightarrow \phi_c$ for these systems can be calculated. Though both $\rho k_BT \kappa_T$ and $S(0)$ approach zero as $\phi \rightarrow\phi_c$, $X$ still diverges with a pole of order one at $\phi = \phi_c$. This is an interesting result, and it strongly indicates that the jammed glassy state for this model is fundamentally nonequilibrium in nature, though at this time it is unclear precisely how $X$ will behave near $\phi_c$ for other glassy states and hyperuniform systems.

We have shown that for a variety of monodisperse hard-sphere systems compressed at various rates from various initial configurations that there is a precursor to the jammed glassy state evident far below the jamming packing fraction $\phi_c$. This precursor appears as linear behavior in the structure factor $S(k)$ near $k=0$ associated with a quickly growing lengthscale $\xi_{SF}$, and for hyperuniform systems is indicative of the onset of an inverted critical point at $\phi = \phi_c$ associated with a diverging length scale $\xi_{DCF}$ and long-ranged power law decay of the direct correlation function $c(r)$. Our identification of these length scales is a unique finding that demonstrates the onset of the jammed glassy state is detectable using standard pair information \cite{ZT2011a}.

Due to the early onset and robustness of the jamming precursor and the dynamics employed in our model of a prototypical glass, we expect that our results are broadly applicable to glass forming molecular systems ({\it e.g.}, metallic glasses, network glasses, etc.). Hard-sphere packings near jamming are known to provide excellent structural models of glassy molecular states \cite{CLPCMP1995,TorquatoRHM2002,PiazzaSM2011}. Endowed with Newtonian dynamics and quickly compressed, hard-sphere models are driven by strong pair repulsion and free-volume dynamics, which are the salient drivers in their molecular counterparts. These drivers lead to diverging elastic moduli and timescales just as sudden macroscopic rigidity and diverging timescales appear in molecular glass formers. Therefore, the precursor to the jammed glassy state and the diverging length scales that we have shown to be present in hard-sphere glasses are applicable broadly to other glass formers \cite{endnote5}; preliminary results indicate that $\xi_{DCF}$ and $\xi_{SF}$ grow as a function of inverse temperature in a supercooled model molecular glass former.

Despite the presence of diverging length scales, we have shown via a nonequilibrium index $X$ that hard-sphere glasses are intrinsically nonequilibrium in nature, where $X$ appears to diverge as $\phi\rightarrow\phi_c$. Though it is currently unclear if $X$ will diverge similarly in supercooled model molecular glass formers, we expect that it will generally increase with decreases in temperature as glassy states are approached. In future work, we will also investigate the behavior of $X$ as a function of temperature in supercooled model molecular glasses.

This work was supported by the NSF under award number DMR-0820341.

%\bibliography{glassyBib1}

\begin{thebibliography}{32}
\expandafter\ifx\csname natexlab\endcsname\relax\def\natexlab#1{#1}\fi
\expandafter\ifx\csname bibnamefont\endcsname\relax
  \def\bibnamefont#1{#1}\fi
\expandafter\ifx\csname bibfnamefont\endcsname\relax
  \def\bibfnamefont#1{#1}\fi
\expandafter\ifx\csname citenamefont\endcsname\relax
  \def\citenamefont#1{#1}\fi
\expandafter\ifx\csname url\endcsname\relax
  \def\url#1{\texttt{#1}}\fi
\expandafter\ifx\csname urlprefix\endcsname\relax\def\urlprefix{URL }\fi
\providecommand{\bibinfo}[2]{#2}
\providecommand{\eprint}[2][]{\url{#2}}

\bibitem[{\citenamefont{Chaikin and Lubensky}(1995)}]{CLPCMP1995}
\bibinfo{author}{\bibfnamefont{P.~M.} \bibnamefont{Chaikin}} \bibnamefont{and}
  \bibinfo{author}{\bibfnamefont{T.~C.} \bibnamefont{Lubensky}},
  \emph{\bibinfo{title}{Principles of Condensed Matter Physics}}
  (\bibinfo{publisher}{Cambridge}, \bibinfo{year}{1995}).

\bibitem[{\citenamefont{Berthier et~al.}(2007)\citenamefont{Berthier, Biroli,
  Bouchaud, Kob, Miyazaki, and Reichman}}]{BBBKMR2007a}
\bibinfo{author}{\bibfnamefont{L.}~\bibnamefont{Berthier}},
  \bibinfo{author}{\bibfnamefont{G.}~\bibnamefont{Biroli}},
  \bibinfo{author}{\bibfnamefont{J.-P.} \bibnamefont{Bouchaud}},
  \bibinfo{author}{\bibfnamefont{J.~W.} \bibnamefont{Kob}},
  \bibinfo{author}{\bibfnamefont{K.}~\bibnamefont{Miyazaki}}, \bibnamefont{and}
  \bibinfo{author}{\bibfnamefont{D.~R.} \bibnamefont{Reichman}},
  \bibinfo{journal}{J. Phys. Chem.} \textbf{\bibinfo{volume}{126}},
  \bibinfo{pages}{184503} (\bibinfo{year}{2007}).

\bibitem[{\citenamefont{Karmakar et~al.}(2009)\citenamefont{Karmakar, Dasgupta,
  and Sastry}}]{KDS2009a}
\bibinfo{author}{\bibfnamefont{S.}~\bibnamefont{Karmakar}},
  \bibinfo{author}{\bibfnamefont{C.}~\bibnamefont{Dasgupta}}, \bibnamefont{and}
  \bibinfo{author}{\bibfnamefont{S.}~\bibnamefont{Sastry}},
  \bibinfo{journal}{PNAS US} \textbf{\bibinfo{volume}{106}},
  \bibinfo{pages}{3675} (\bibinfo{year}{2009}).

\bibitem[{\citenamefont{Chandler and Garrahan}(2010)}]{CG2010a}
\bibinfo{author}{\bibfnamefont{D.}~\bibnamefont{Chandler}} \bibnamefont{and}
  \bibinfo{author}{\bibfnamefont{J.~P.} \bibnamefont{Garrahan}},
  \bibinfo{journal}{Ann. Rev. Phys. Chem.} \textbf{\bibinfo{volume}{61}},
  \bibinfo{pages}{191} (\bibinfo{year}{2010}).

\bibitem[{\citenamefont{Lubchenko and Wolynes}(2006)}]{LW2006a}
\bibinfo{author}{\bibfnamefont{V.}~\bibnamefont{Lubchenko}} \bibnamefont{and}
  \bibinfo{author}{\bibfnamefont{P.~G.} \bibnamefont{Wolynes}},
  \bibinfo{journal}{Ann. Rev. Phys. Chem.} \textbf{\bibinfo{volume}{58}},
  \bibinfo{pages}{235} (\bibinfo{year}{2006}).

\bibitem[{\citenamefont{Hocky et~al.}(2012)\citenamefont{Hocky, Markland, and
  Reichman}}]{HMR2012a}
\bibinfo{author}{\bibfnamefont{G.~M.} \bibnamefont{Hocky}},
  \bibinfo{author}{\bibfnamefont{T.~E.} \bibnamefont{Markland}},
  \bibnamefont{and} \bibinfo{author}{\bibfnamefont{D.~R.}
  \bibnamefont{Reichman}} (\bibinfo{year}{2012}), \eprint{arXiv:1201.2888}.

\bibitem[{\citenamefont{Torquato et~al.}(2000)\citenamefont{Torquato, Truskett,
  and Debenedetti}}]{TTD2000a}
\bibinfo{author}{\bibfnamefont{S.}~\bibnamefont{Torquato}},
  \bibinfo{author}{\bibfnamefont{T.~M.} \bibnamefont{Truskett}},
  \bibnamefont{and} \bibinfo{author}{\bibfnamefont{P.~G.}
  \bibnamefont{Debenedetti}}, \bibinfo{journal}{Phys. Rev. Lett.}
  \textbf{\bibinfo{volume}{84}}, \bibinfo{pages}{2064} (\bibinfo{year}{2000}).

\bibitem[{\citenamefont{Torquato and Stillinger}(2007)}]{TS2007a}
\bibinfo{author}{\bibfnamefont{S.}~\bibnamefont{Torquato}} \bibnamefont{and}
  \bibinfo{author}{\bibfnamefont{F.~H.} \bibnamefont{Stillinger}},
  \bibinfo{journal}{J. App. Phys.} \textbf{\bibinfo{volume}{102}},
  \bibinfo{pages}{093511} (\bibinfo{year}{2007}); ibid
  \textbf{\bibinfo{volume}{103}},
  \bibinfo{pages}{129902} (\bibinfo{year}{2008}).

\bibitem[{\citenamefont{Torquato and Stillinger}(2010)}]{TS2010a}
\bibinfo{author}{\bibfnamefont{S.}~\bibnamefont{Torquato}} \bibnamefont{and}
  \bibinfo{author}{\bibfnamefont{F.~H.} \bibnamefont{Stillinger}},
  \bibinfo{journal}{Rev. Mod. Phys.} \textbf{\bibinfo{volume}{82}},
  \bibinfo{pages}{2633} (\bibinfo{year}{2010}).

\bibitem[{\citenamefont{Torquato and Stillinger}(2003)}]{TS2003a}
\bibinfo{author}{\bibfnamefont{S.}~\bibnamefont{Torquato}} \bibnamefont{and}
  \bibinfo{author}{\bibfnamefont{F.~H.} \bibnamefont{Stillinger}},
  \bibinfo{journal}{Phys. Rev. E} \textbf{\bibinfo{volume}{68}},
  \bibinfo{pages}{041113} (\bibinfo{year}{2003}).

\bibitem[{\citenamefont{Donev et~al.}(2005{\natexlab{a}})\citenamefont{Donev,
  Stillinger, and Torquato}}]{DST2005a}
\bibinfo{author}{\bibfnamefont{A.}~\bibnamefont{Donev}},
  \bibinfo{author}{\bibfnamefont{F.~H.} \bibnamefont{Stillinger}},
  \bibnamefont{and} \bibinfo{author}{\bibfnamefont{S.}~\bibnamefont{Torquato}},
  \bibinfo{journal}{Phys. Rev. Lett.} \textbf{\bibinfo{volume}{95}},
  \bibinfo{pages}{090604} (\bibinfo{year}{2005}{\natexlab{a}}).

\bibitem[{end({\natexlab{a}})}]{endnote1}
\bibinfo{note}{For a hard-sphere system, compression qualitatively plays the
  same role as decreasing the temperature in an atomic or molecular system; see
  Ref. \cite{TS2010a}.}

\bibitem[{\citenamefont{Reatto and Chester}(1967)}]{RC1967a}
\bibinfo{author}{\bibfnamefont{L.}~\bibnamefont{Reatto}} \bibnamefont{and}
  \bibinfo{author}{\bibfnamefont{G.~V.} \bibnamefont{Chester}},
  \bibinfo{journal}{Phys. Rev.} \textbf{\bibinfo{volume}{155}},
  \bibinfo{pages}{88} (\bibinfo{year}{1967}).

\bibitem[{\citenamefont{Torquato et~al.}(2008)\citenamefont{Torquato,
  Scardicchio, and Zachary}}]{TSZ2008a}
\bibinfo{author}{\bibfnamefont{S.}~\bibnamefont{Torquato}},
  \bibinfo{author}{\bibfnamefont{A.}~\bibnamefont{Scardicchio}},
  \bibnamefont{and} \bibinfo{author}{\bibfnamefont{C.}~\bibnamefont{Zachary}},
  \bibinfo{journal}{J. Stat. Mech., Theor. Exp.} p. \bibinfo{pages}{P11019}
  (\bibinfo{year}{2008}).

\bibitem[{\citenamefont{Peebles}(1993)}]{PeeblesPC1993}
\bibinfo{author}{\bibfnamefont{P.~J.~E.}~\bibnamefont{Peebles}},
  \emph{\bibinfo{title}{Principles of Cosmology}}
  (\bibinfo{publisher}{Princeton University Press}, \bibinfo{year}{1993}).

\bibitem[{\citenamefont{Jiao and Torquato}(2011)}]{JT2011a}
\bibinfo{author}{\bibfnamefont{Y.}~\bibnamefont{Jiao}} \bibnamefont{and}
  \bibinfo{author}{\bibfnamefont{S.}~\bibnamefont{Torquato}},
  \bibinfo{journal}{Phys. Rev. E} \textbf{\bibinfo{volume}{84}},
  \bibinfo{pages}{041309} (\bibinfo{year}{2011}).

\bibitem[{\citenamefont{Long et~al.}(March 11th-15th, 2012)\citenamefont{Long,
  Xie, Weigand, Moss, Roorda, Torquato, and Steinhardt}}]{LXWMRTS2012a}
\bibinfo{author}{\bibfnamefont{G.}~\bibnamefont{Long}},
  \bibinfo{author}{\bibfnamefont{R.}~\bibnamefont{Xie}},
  \bibinfo{author}{\bibfnamefont{S.}~\bibnamefont{Weigand}},
  \bibinfo{author}{\bibfnamefont{S.}~\bibnamefont{Moss}},
  \bibinfo{author}{\bibfnamefont{S.}~\bibnamefont{Roorda}},
  \bibinfo{author}{\bibfnamefont{S.}~\bibnamefont{Torquato}}, \bibnamefont{and}
  \bibinfo{author}{\bibfnamefont{P.}~\bibnamefont{Steinhardt}},
  \bibinfo{journal}{Proceedings of the Minerals, Metals and Materials Society}
  (\bibinfo{year}{March 11th-15th, 2012}), \bibinfo{note}{Orlando, FL (to be
  published)}.

\bibitem[{\citenamefont{Donev et~al.}(2005{\natexlab{b}})\citenamefont{Donev,
  Torquato, and Stillinger}}]{DTS2005a}
\bibinfo{author}{\bibfnamefont{A.}~\bibnamefont{Donev}},
  \bibinfo{author}{\bibfnamefont{S.}~\bibnamefont{Torquato}}, \bibnamefont{and}
  \bibinfo{author}{\bibfnamefont{F.~H.} \bibnamefont{Stillinger}},
  \bibinfo{journal}{J. Comp. Phys.} \textbf{\bibinfo{volume}{202}},
  \bibinfo{pages}{737} (\bibinfo{year}{2005}{\natexlab{b}}).

\bibitem[{end({\natexlab{b}})}]{endnote3}
\bibinfo{note}{Fits are generated by least-squares over the range $0.05 \leq k
  \leq 0.25$ employing weights equivalent to the inverse of the variance over
  the $20$ packings of the values of $S(k)$ at each $k$. The range $0.05 \leq k
  \leq 0.25$ captures the behavior of $S(k)$ closest to the origin while
  excluding the smallest values of $k$, which, as a result of the finite size
  of the packings, exhibit substantial volatility due to the small numbers of
  points available to calculate $S(k)$. We note, however, that the inclusion of
  the points for $k < 0.05$ does not significantly alter the fits or their
  R-squared values.}

\bibitem[{\citenamefont{Torquato}(2002)}]{TorquatoRHM2002}
\bibinfo{author}{\bibfnamefont{S.}~\bibnamefont{Torquato}},
  \emph{\bibinfo{title}{Random Heterogeneous Materials}}
  (\bibinfo{publisher}{Springer}, \bibinfo{year}{2002}).

\bibitem[{\citenamefont{Hansen and McDonald}(2006)}]{HMTSL2006}
\bibinfo{author}{\bibfnamefont{J.~P.} \bibnamefont{Hansen}} \bibnamefont{and}
  \bibinfo{author}{\bibfnamefont{I.~R.} \bibnamefont{McDonald}},
  \emph{\bibinfo{title}{Theory of Simple Liquids, 3rd ed.}}
  (\bibinfo{publisher}{Academic}, \bibinfo{year}{2006}).

\bibitem[{\citenamefont{Levesque et~al.}(1999)\citenamefont{Levesque, Weis, and
  Lebowitz}}]{LWL1999a}
\bibinfo{author}{\bibfnamefont{D.}~\bibnamefont{Levesque}},
  \bibinfo{author}{\bibfnamefont{J.~J.} \bibnamefont{Weis}}, \bibnamefont{and}
  \bibinfo{author}{\bibfnamefont{J.~L.} \bibnamefont{Lebowitz}},
  \bibinfo{journal}{J. Stat. Phys.} \textbf{\bibinfo{volume}{100}},
  \bibinfo{pages}{209} (\bibinfo{year}{1999}).

\bibitem[{HST(2012)}]{HST2012a}
	\bibinfo{author}{\bibfnamefont{A.~B.} \bibnamefont{Hopkins}},
	\bibinfo{author}{\bibfnamefont{F.~H.} \bibnamefont{Stillinger}}, \bibnamefont{and}
	\bibinfo{author}{\bibfnamefont{S.}~\bibnamefont{Torquato}},
 (\bibinfo{year}{2012}), \bibinfo{note}{manuscript in preparation}.

\bibitem[{\citenamefont{Berthier and Kob}(2011)}]{BK2011a}
\bibinfo{author}{\bibfnamefont{L.}~\bibnamefont{Berthier}} \bibnamefont{and}
  \bibinfo{author}{\bibfnamefont{W.}~\bibnamefont{Kob}},
  \bibinfo{journal}{Phys. Rev. E} \textbf{\bibinfo{volume}{85}},
  \bibinfo{pages}{011102} (\bibinfo{year}{2011}).

\bibitem[{\citenamefont{Charbonneau et~al.}(2012)\citenamefont{Charbonneau,
  Charbonneau, and Tarjus}}]{CCT2012a}
\bibinfo{author}{\bibfnamefont{B.}~\bibnamefont{Charbonneau}},
  \bibinfo{author}{\bibfnamefont{P.}~\bibnamefont{Charbonneau}},
  \bibnamefont{and} \bibinfo{author}{\bibfnamefont{G.}~\bibnamefont{Tarjus}},
  \bibinfo{journal}{Phys. Rev. Lett.} \textbf{\bibinfo{volume}{108}},
  \bibinfo{pages}{035701} (\bibinfo{year}{2012}).

\bibitem[{\citenamefont{Rintoul and Torquato}(1998)}]{RT1998a}
\bibinfo{author}{\bibfnamefont{M.~D.} \bibnamefont{Rintoul}} \bibnamefont{and}
  \bibinfo{author}{\bibfnamefont{S.}~\bibnamefont{Torquato}},
  \bibinfo{journal}{Phys. Rev. E} \textbf{\bibinfo{volume}{58}},
  \bibinfo{pages}{3083} (\bibinfo{year}{1998}).

\bibitem[{\citenamefont{Zachary and Torquato}(2011)}]{ZT2011a}
\bibinfo{author}{\bibfnamefont{C.}~\bibnamefont{Zachary}} \bibnamefont{and}
  \bibinfo{author}{\bibfnamefont{S.}~\bibnamefont{Torquato}},
  \bibinfo{journal}{Phys. Rev. E} \textbf{\bibinfo{volume}{84}},
  \bibinfo{pages}{056102} (\bibinfo{year}{2011}).
  \bibinfo{note}{This paper suggests that generalized two-point cluster functions are more sensitive to long-range structural characteristics in glassy systems than are standard pair correlation functions.}

\bibitem[{\citenamefont{Piazza}(2011)}]{PiazzaSM2011}
\bibinfo{author}{\bibfnamefont{R.}~\bibnamefont{Piazza}},
  \emph{\bibinfo{title}{Soft matter: the stuff that dreams are made of}}
  (\bibinfo{publisher}{Springer}, \bibinfo{year}{2011}).

\bibitem[{end({\natexlab{c}})}]{endnote5}
\bibinfo{note}{The hyperuniformity of maximally random jammed packings has been
  extended to apply to polydisperse spheres and nonspherical objects in terms
  of the spectral density $\tilde{\chi}(k)$, defined as the Fourier transform
  of $\chi(r) = S_2(r) - \phi^2$ with $S_2(r)$ the two-point probability
  function \cite{ZJT2011a}.}

\bibitem[{\citenamefont{Zachary et~al.}(2011)\citenamefont{Zachary, Jiao, and
  Torquato}}]{ZJT2011a}
\bibinfo{author}{\bibfnamefont{C.}~\bibnamefont{Zachary}},
  \bibinfo{author}{\bibfnamefont{Y.}~\bibnamefont{Jiao}}, \bibnamefont{and}
  \bibinfo{author}{\bibfnamefont{S.}~\bibnamefont{Torquato}},
  \bibinfo{journal}{Phys. Rev. Lett.} \textbf{\bibinfo{volume}{106}},
  \bibinfo{pages}{178001} (\bibinfo{year}{2011}).

\end{thebibliography}
%\bibliographystyle{apsrev}

\end{document}